\documentclass[
reprint,
superscriptaddress,
nofootinbib,
amsmath,amssymb,
aps,
pra,
noeprint,
]{revtex4-2}

\usepackage{graphicx}
\usepackage[english]{babel}
\usepackage{appendix}
\usepackage{dcolumn}
\usepackage{bm}
\usepackage{color}
\usepackage{booktabs,eqparbox}
\usepackage{braket}
\usepackage{dsfont}
\usepackage{placeins}
\usepackage{paralist}
\usepackage{comment}
\usepackage{multirow}
\usepackage[utf8]{inputenc}
\usepackage{pgfplots}
\usepackage{amsmath}
\usepackage{graphicx, import}
\usepackage{lipsum}
\usepackage{tikz}
\usepackage{csquotes}
\usetikzlibrary{shapes,arrows,fit, backgrounds, quantikz2}
\usepackage[colorlinks=true, allcolors=magenta]{hyperref}
\usepackage{cleveref}
\usepackage{textgreek}
\usepackage{siunitx}
\usepackage[nolist,nohyperlinks]{acronym}

\DeclareUnicodeCharacter{2212}{−}
\usepgfplotslibrary{groupplots,dateplot}
\usetikzlibrary{patterns,shapes.arrows}
\pgfplotsset{compat=newest}

\makeatletter
\let\oldtheequation\theequation
\renewcommand\tagform@[1]{\maketag@@@{\ignorespaces#1\unskip\@@italiccorr}}
\renewcommand\theequation{(\oldtheequation)}
\makeatother

\definecolor{myRed}{HTML}{A3061E}
\definecolor{myBlue} {RGB} {0,63,119}
\definecolor{myYellow} {cmy} {0,0.263,0.741}
\definecolor{myGreen}{HTML}{0B6E4F}
\definecolor{FAU}{RGB}{0,56,101}
\definecolor{gray75}{gray}{0.75}
\colorlet{myOrange} {myYellow!60!myRed}
\colorlet{myViolet}{myRed!50!myBlue!80}

\renewcommand{\vec}[1]{\bm{#1}}

\newcommand{\lp}{\left(}
\newcommand{\rp}{\right)}
\newcommand{\lc}{\left\{}
\newcommand{\rc}{\right\}}

\addto\extrasenglish{
  
}

\addto\extrasenglish{
  
}

\graphicspath{{Figures/}}

\begin{document}
\title{Solving an Industrially Relevant Quantum Chemistry Problem on Quantum Hardware}

\author{Ludwig Nützel }
\email{Ludwig.Nuetzel@fau.de}
\affiliation{Department of Physics, Friedrich-Alexander Universität Erlangen-Nürnberg, Erlangen, Germany}

\author{Alexander Gresch }
\affiliation{Heinrich Heine University Düsseldorf, Faculty of Mathematics and Natural Sciences, Germany}
\affiliation{Institute for Quantum Inspired and Quantum Optimization, Hamburg University of Technology, Germany}

\author{Lukas Hehn }
\affiliation{BASF SE, Next Generation Computing, Pfalzgrafenstr.~1, 67061 Ludwigshafen, Germany}

\author{Lucas Marti }
\affiliation{Department of Physics, Friedrich-Alexander Universität Erlangen-Nürnberg, Erlangen, Germany}

\author{Robert Freund}
\affiliation{Universit{\"a}t Innsbruck, Institut f{\"u}r Experimentalphysik, Innsbruck, Austria}

\author{Alex Steiner}
\affiliation{Universit{\"a}t Innsbruck, Institut f{\"u}r Experimentalphysik, Innsbruck, Austria}

\author{Christian D. Marciniak}
\affiliation{Universit{\"a}t Innsbruck, Institut f{\"u}r Experimentalphysik, Innsbruck, Austria}

\author{Timo Eckstein }
\affiliation{Department of Physics, Friedrich-Alexander Universität Erlangen-Nürnberg, Erlangen, Germany}
\affiliation{Max Planck Institute for the Science of Light, Staudtstraße 2, 91058 Erlangen, Germany}

\author{Nina Stockinger }
\affiliation{Department of Physics, Friedrich-Alexander Universität Erlangen-Nürnberg, Erlangen, Germany}
\affiliation{Department of Chemistry, Technical University of Munich, Garching, Germany}

\author{Stefan Wolf }
\affiliation{Department of Physics, Friedrich-Alexander Universität Erlangen-Nürnberg, Erlangen, Germany}

\author{Thomas Monz}
    \altaffiliation[Also at ]{Alpine Quantum Technologies GmbH, Innsbruck, Austria}
\affiliation{Universit{\"a}t Innsbruck, Institut f{\"u}r Experimentalphysik, Innsbruck, Austria}

\author{Michael Kühn }
\affiliation{BASF SE, Next Generation Computing, Pfalzgrafenstr.~1, 67061 Ludwigshafen, Germany}

\author{Michael J. Hartmann }
\affiliation{Department of Physics, Friedrich-Alexander Universität Erlangen-Nürnberg, Erlangen, Germany}
\affiliation{Max Planck Institute for the Science of Light, Staudtstraße 2, 91058 Erlangen, Germany}

    \begin{abstract}
       {
        Quantum chemical calculations are among the most promising applications for quantum computing. Implementations of dedicated quantum algorithms on available quantum hardware were so far, however, mostly limited to comparatively simple systems without strong correlations. As such, they can also be addressed by classically efficient single-reference methods. In this work, we calculate the lowest energy eigenvalue of active space Hamiltonians of industrially relevant and strongly correlated metal chelates on trapped ion quantum hardware, and integrate the results into a typical industrial quantum chemical workflow to arrive at chemically meaningful properties. 
We are able to achieve chemical accuracy by training a variational quantum algorithm on quantum hardware, followed by a classical diagonalization in the subspace of states measured as outputs of the quantum circuit. This approach is particularly measurement-efficient, requiring 600 single-shot measurements per cost function evaluation on a ten qubit system, and allows for efficient post-processing to handle erroneous runs.
       }
    \end{abstract}

\maketitle

\section{Introduction}
Molecular quantum chemistry is one of the most promising fields for quantum computing applications due to the exponential scaling of classical simulation cost with the number of molecular orbitals~\cite{cao2019,mcardle2020}.
This makes exact classical simulations impractical for all but the smallest molecules. Quantum algorithms for solving the electronic structure problem, in turn, are expected to scale polynomially with system size and accuracy requirements~\cite{bauer2020,motta2022}.
Even though quantum computing may not guarantee exponential advantage in estimating a quantum chemical system's \ac{GSE}~\cite{lee2023}, the complexity of these problems~\cite{Bookatz_2013,Cade_Weggemans_2023} strongly motivates the pursuit of quantum algorithms, which appear to be more promising than mere enhancements to classical methods~\cite{cao2019}.
This is especially the case in strongly correlated systems where classical approximations, such as \ac{DFT}~\cite{Kohn1965} and coupled cluster methods like CCSD(T)~\cite{cizek1966}, often fail to provide the required accuracy. For these systems, the use of multi-reference active space methods such as \ac{CASCI} and \ac{CASSCF} is typically needed to appropriately account for the strong electron correlation~\cite{Roca2012}. However, the exponential increase in the classical simulation cost with the active space size motivates research into solving such active spaces on quantum hardware~\cite{Liepuoniute2024EF,Ollitrault2024,obrien2023,Motta2023EF,elfving2020,Otten2022,Jensen2024}.

\begin{figure}[t!]
    \centering
    \includegraphics[width=0.7\linewidth]{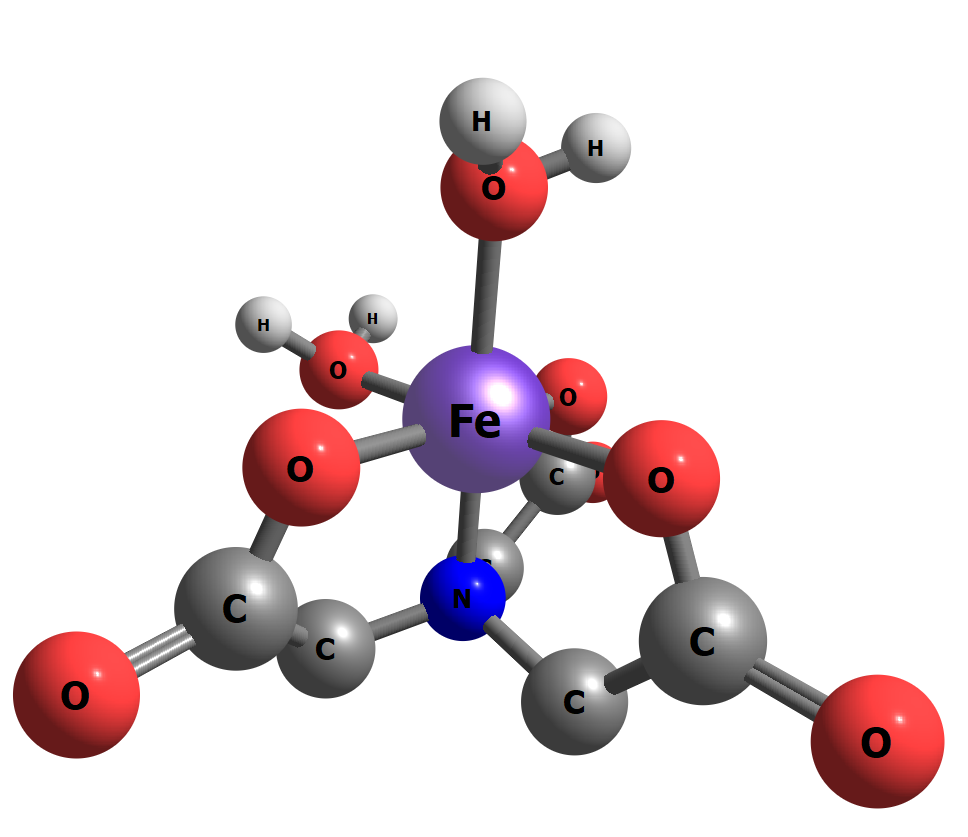}
\caption{{\bf{Geometry of Fe(III)-NTA.}} The geometry of the Fe(III)-NTA complex $\rm [Fe (NTA) (H_2O)_2]$ in its low-spin state is shown. Six hydrogen atoms are hidden for visual clarity.}
\label{fig:geo_NTA}
\end{figure}

So far, most of the quantum chemical calculations on quantum hardware focus on molecules that do not display strong electron correlations or are not industrially relevant~\cite{guo2024,sun2024evaluating,Liepuoniute2024EF,obrien2023,Motta2023EF,Zhao2023,khan2023,huggins2022AFQMC,Eddins2022EF,rice2021,Arute2020,Nam2020,kandala2019,McCaskey2019,Colless2018,Kandala2017,OMalley2016,Peruzzo2014}.
In this work, we use quantum hardware to solve active space Hamiltonians of the industrially relevant and strongly correlated Fe(III)-NTA metal chelate.
Fe(III)-NTA consists of NTA, the trianion formed by deprotonating the chelating agent nitrilotriacetic acid, coordinating to an Fe(III) ion as shown in \autoref{fig:geo_NTA}.
Chelating agents are used in household and industrial applications for a variety of purposes, for example as water softeners in cleaning applications, as ligands for catalysts, as antioxidants, and in soil remediation~\cite{checa2021ChelatingApp,gulcin2022ChelatingApp,prete2021ChelatingWater,eivazihollagh2019chelatingApp}.Accurate simulations are needed to enhance the development of new chelating agents, as they in principle allow to preselect the most promising candidate molecules. Thus, the amount of time-consuming and expensive laboratory trials can be reduced.

For transition metal compounds like Fe(III)-NTA,
an accurate calculation of the energetic separation between different spin states
is generally needed to derive spectroscopic properties and
subsequently determine the energetically most favorable spin state, which is the starting point for calculating properties of such compounds and respective chemical reactions~\cite{Verma2017, radon2015}.
Thus, an important step in the simulation of the prototypical Fe(III)-NTA chelate complex is calculating the separation between the energetically close-lying doublet (\ac{LS}), quartet (\ac{IS}), and sextet (\ac{HS}) states. This
poses a significant challenge for electronic structure methods, and is an application where multi-reference active space methods can be advantageous over most single-reference methods~\cite{Hehn2024}.

Here, we use a novel variational quantum algorithm enhanced with subspace diagonalization as classical post-processing to determine the \acp{GSE} of the \ac{IS} and \ac{LS} active space Hamiltonians of Fe(III)-NTA on trapped ion quantum hardware, wherein the active space Hamiltonian of the \ac{LS} state represents one of the most complex quantum chemical Hamiltonians solved on quantum hardware to date.

\begin{figure*}[t!]
    \centering
    \scalebox{0.9}{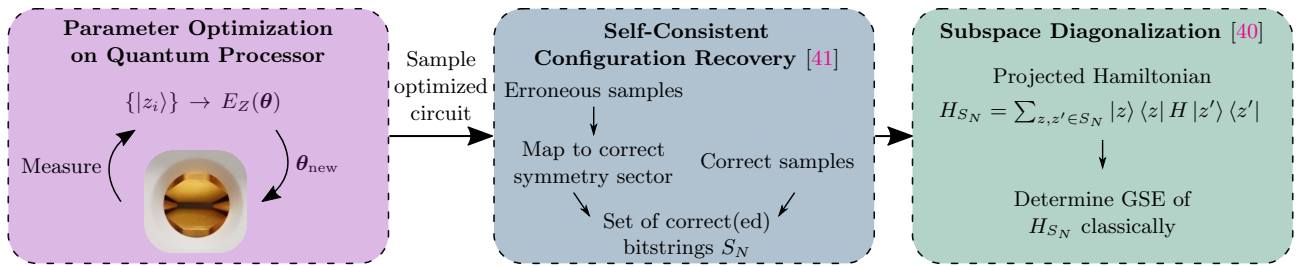}
    \caption{
        \label{fig:workflow}\textbf{Workflow summary.} \textit{Left}: Optimization of quantum circuit parameters. Sets of parameters are fed to the quantum processor, e.g. a trapped ion quantum processor (bottom), measurements in the computational basis allow the estimation of the diagonal terms $E_Z$ of $H$, and this expectation value is minimized.
        \textit{Middle:} After the optimization has finished, the final circuit is sampled in the computational basis. This yields two types of bitstrings, (i) those that have the correct particle number, and (ii) those that do not.
        A self-consistent configuration recovery scheme~\cite{RobledoMoreno2024} is used to map the erroneous bitstrings back to the correct symmetry sector.
        \textit{Right:} The problem Hamiltonian $H$ is reduced to the subspace of the correct(ed) bitstrings, obtained in the previous step, via projection.
        Classical numerically exact diagonalization of the projected Hamiltonian $H_{S_N}$ provides an upper bound to the exact \acf{GSE}.
    }
\end{figure*}

Our approach starts with identifying an active space from a \ac{HF} calculation and computing the necessary integrals to express the respective fermionic Hamiltonian in second quantization. This second-quantized Hamiltonian is then mapped to the Pauli basis using the Jordan-Wigner encoding.
To find the \acp{GSE} of this Hamiltonian, we train a parameterized quantum circuit that conserves the particle number in each spin sector on a trapped ion quantum computer to prepare quantum states within a relevant subspace of the problem's Hilbert space. Diagonalizing the Hamiltonian restricted to this subspace during classical post-processing, we then find its \ac{GSE} for the selected spin state.
We use further classical post-processing techniques on the measurement outcomes to enhance the accuracy of the hardware results.
Finally, the obtained \acp{GSE} for the \ac{LS} and \ac{IS} states are used to calculate the relative Gibbs free energy of the two spin states in aqueous solution, which is the target thermodynamic property of interest to chemists.

\section{Quantum Algorithm\label{sec:vqe}}
To find the \ac{GSE} of an electronic structure Hamiltonian, we designed a quantum algorithm that is inspired by the \acf{VQE}~\cite{Peruzzo2014,McClean2016,Bittel_2022,Zhenyu2020,Stilck_Franca_2021,Cerezo2021}.
The \ac{VQE} is a hybrid quantum-classical algorithm that minimizes an estimate of the \ac{GSE} via a trial state $\ket{\psi_{\Vec{\theta}}}$, that is obtained from an initial state $\ket{\psi_0}$ via a parameterized quantum circuit $C(\Vec{\theta})$, $\ket{\psi_{\Vec{\theta}}}=C(\Vec{\theta})\ket{\psi_0}$.
The expectation value of this trial state sets an upper bound to the \ac{GSE} $E_0$.
The \ac{VQE} minimizes this upper bound by optimizing the circuit parameters $\Vec{\theta}$.
The quality of this estimate depends on the expressibility and trainability of the quantum circuit, the energy estimation from repeated state preparations, and the parameter update scheme.
Expressive quantum circuits often suffer from the barren plateau problem, where the gradients of the cost function, and variances thereof, vanish exponentially with the number of qubits, thus requiring exponential resources for optimization ~\cite{McClean2018,Cerezo2021}. As a consequence, \ac{VQE} applications require large amounts of measurements~\cite{Gonthier2022MeasurementBottleneck}, since all individual terms in the Pauli decomposition of the Hamiltonian $H$ need to be estimated with very high precision to resolve the gradients.

To ease these challenges of \ac{VQE} algorithms, we have modified the parameter optimization procedure and added a classical post-processing step. Our VQE-inspired algorithm consists of the following three ingredients, which are also summarized in \autoref{fig:workflow}.
First, a variational quantum algorithm prepares a quantum state in a reduced subspace of the Hilbert space to yield good approximations of the desired ground state. This step proceeds in analogy to a \ac{VQE} algorithm, but instead of estimating the expectation value $\bra{\psi_{\Vec{\theta}}} H \ket{\psi_{\Vec{\theta}}}$, we only measure the prepared state in the computational basis and use
\begin{align} \label{eq:comp_base_estimate}
    E_Z(\Vec{\theta}) = \frac{1}{N_\mathrm{tot}} \sum_{j=1}^{N_{u}} N_j \bra{z_j} H \ket{z_j},
\end{align}
as a cost function for updating the variational parameters $\Vec{\theta}$. Here $\ket{z_j}$ is a measured computational basis state, $N_u$ the number of unique measured bitstrings, $N_j$ the number of times $\ket{z_j}$ was measured, and $N_\mathrm{tot} = \sum_{j=1}^{N_u} N_j$ the total number of measurements. The choice of restricting measurements to the Pauli-$Z$ basis is motivated by the fact that, for our model,  this allows for efficient post-selection to discard faulty circuit executions, as discussed below, as well as the dominating diagonal terms in $H$. In other applications, measurements in several different bases might be preferable~\cite{Gresch2023ShadowGrouping}.

\autoref{eq:comp_base_estimate} is a biased estimate as we do not measure all the terms contained in $H$.
Hence, we rectify this for the final read-out as follows.
We once again sample the final prepared state in the computational basis and subsequently recover erroneous bitstrings via \ac{SCCR}~\cite{RobledoMoreno2024}.
Finally, we diagonalize the problem Hamiltonian $H$, projected on this recovered subspace, in classical post-processing~\cite{Kanno2023}. This is always feasible since the number of measured bitstrings and hence the dimension of the recovered subspace is limited by the total measurement budget $N_\mathrm{tot}$.
This furthermore circumvents the need of having to measure the many other terms of $H$.

\subsection{Problem Structure and variational gate sequence}
The quantum part of our approach considers the electronic structure Hamiltonian of the active space,
\begin{align}
    \label{eq:electronic_structure_ham}
    H_\mathrm{el.} = \sum_{ij,\sigma} t_{ij}^{\sigma} c_{i\sigma}^\dagger c_{j\sigma}^{\phantom{\dagger}}  + \sum_{ijkl,\sigma\sigma^\prime} t_{ijkl}^{\sigma\sigma^\prime} c_{i\sigma}^\dagger c_{j\sigma^\prime}^\dagger c_{k\sigma}^{\phantom{\dagger}} c_{l\sigma^\prime}^{\phantom{\dagger}} \,,
\end{align}
where $c^\dagger$ and $c$ are fermionic creation and annihilation operators, $\{i,j,k,l\}$ index spatial orbitals, $\sigma \in \{\uparrow, \downarrow\}$ denotes the spin, and $t_{ij}$ and $t_{ijkl}$ are the one- and two-body integrals in the basis that is given by the molecular orbitals obtained from a \ac{CASSCF} calculation on classical hardware.
These Hamiltonians conserve particle numbers in each spin sector by construction.
Therefore, a \ac{NP} circuit is a natural choice, as it restricts the Hilbert space explored by the circuit to states with the correct particle numbers.
We further choose to map the electronic structure Hamiltonian in \autoref{eq:electronic_structure_ham} to a qubit Hamiltonian via the Jordan-Wigner encoding.

\begin{figure}[ht!]
    \centering
    \includegraphics[width=0.85\linewidth]{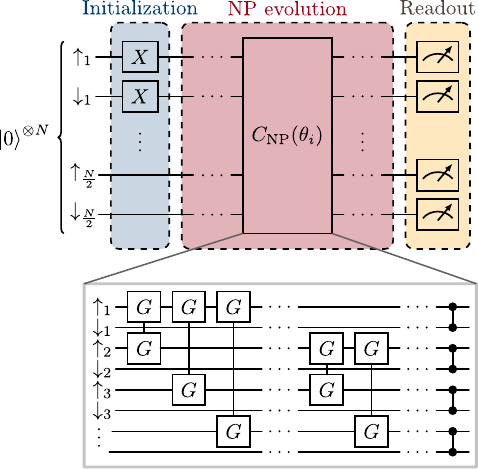}
    \caption{\textbf{Schematic of the circuit ansatz.} The circuit is initialized to the \ac{HF} state, evolved using an \ac{NP} ansatz, and finally read out in the computational basis.
    The inset shows one layer of the variational ansatz $C_\mathrm{NP}(\Vec{\theta}_i)$. Number-preserving Givens rotations (see Appendix~\ref{subsec:circuit_ansatz} \autoref{eq:givens} for a definition) within the spin sectors evolve the input state, subsequent CZ gates entangle the spin sectors without exchanging particles. The rotations displayed for the spin-up sector are simultaneously applied to the spin-down sector.\label{fig:circuit_ansatz}}
\end{figure}

In addition to particle number, the electronic structure Hamiltonian also conserves the total spin $S^2$~\cite{Anselmetti2021},
\begin{align}
    S^2 = \sum_{i,j=1}^{N/2} c_{i,\uparrow} c_{j,\uparrow}^\dagger c_{i,\downarrow}^\dagger c_{j,\downarrow}
    + \frac{n_\uparrow - n_\downarrow}{2} + \frac{(n_\uparrow - n_\downarrow)^2}{4}\,.
    \label{eq:total_spin_squared}
\end{align}
Finding ground states with the correct total spin is necessary to calculate the energetic separation between two different spin states in a subsequent step.
This can e.g.\ be achieved by initializing the circuit with the \ac{HF} state, which has the correct spin squared value, and then evolving the state using \ac{QNP} gates~\cite{Anselmetti2021}.
However, decomposing \ac{QNP} gates into two-qubit gate equivalents (2QGE), which in our case corresponds to $R_{XX}$ gates, is costly, and the benefits of using a \ac{QNP} ansatz are quickly overshadowed by the limitations of current noisy intermediate-scale quantum devices.
We therefore propose a new \ac{NP} ansatz, shown in \autoref{fig:circuit_ansatz}
and detailed in Appendix~\ref{subsec:circuit_ansatz}, that conserves the particle number per spin sector, but does not conserve $S^2$.
The ansatz follows a layered structure and can be deepened to fit the problem at hand.
To enforce a correct value of $S^2$, we add a penalty term to the Hamiltonian,
\begin{align}
     H = H_\mathrm{el.} + M\lp s_0^2 -  S^2 \rp^2,
     \label{eq:penalized_hamiltonian}
\end{align}
where $s_0^2$ is the desired total spin squared value that takes on the value $s_0^2=15/4\ (3/4)$ for the \ac{IS} (\ac{LS}) state of Fe(III)-NTA.
The penalty strength $M>0$ is a scalar that changes the spectrum such that the \ac{GSE} of $H_\mathrm{el.}$, $E_{0}$, is raised above the \ac{GSE} of a spin configuration with total spin squared $s_0^2$.
The ground state of the desired spin configuration becomes the ground state of $H$ with eigenvalue $E_{0,s_0^2}$.
For this work we choose a value of \mbox{$M=0.01$}.

\subsection{Circuit read-out and post-processing}
\label{subsec:read_out}
After preparing a trial state on the quantum device, we need to measure the cost function
$E_Z(\Vec{\theta})$ in \autoref{eq:comp_base_estimate}.
Here, our strategy to only measure in the computational basis (the $Z^{\otimes N}$-Pauli basis) enables us to use the following, very efficient post-selection techniques.

\paragraph{Post-selection:}
Since we only measure $Z$-Pauli operators, we can immediately check for a possible break of the underlying symmetry of the system.
We will use this post-selection technique repeatedly during the parameter optimization on quantum hardware.
Upon measurement, the trial state is projected onto a computational basis state.
The circuit is number-preserving, meaning that every trial state is a superposition of computational basis states with the same particle number as the initial state.
For a measured computational basis state $\ket{z_m}$ this yields the following constraint,
\begin{align}
    \sum_{i=1}^{N/2} \bra{\psi_\mathrm{HF}}  Z_{i,\sigma} \ket{\psi_\mathrm{HF}} = \sum_{i=1}^{N/2} \bra{z_{\mathrm{m}}}  Z_{i,\sigma} \ket{z_{\mathrm{m}}}, \label{eq:postselection_constraint}
\end{align}
where $\ket{\psi_\mathrm{HF}}$ is the \ac{HF} (i.e. initial) state and $\sigma \in \lc \uparrow, \downarrow \rc$.
If the measured computational basis state has a wrong particle number in at least one of the two spin sectors, we know that an error has occurred and can immediately discard this run.

\paragraph{\ac{SCCR}:}
To map erroneous runs back to the correct symmetry sector, we use the method of Ref.~\cite{RobledoMoreno2024} on the final samples obtained after the parameter optimization has converged.
To this end, the sampled bitstrings are categorized into a correct set $S_N$ and an incorrect set $S_X$ using the post-selection described above.
At first, $S_N$ is split into $K$ batches labeled $S_N^{(k)}$.
Each batch contains bitstrings $z_i^{k}$ that span a subspace $\{\ket{z_i^{k}}\}$.
This gives rise to a projected Hamiltonian
\begin{align}
    H_{S_N^{(k)}} = P_{S_N^{(k)}} H P_{S_N^{(k)}},
\end{align}
where $P_{S_N^{(k)}} = \sum_{z_i^{k} \in S_N^{(k)}} \ket{z_i^{k}} \bra{z_i^{k}}$.
The batch sizes grow polynomially with system size, allowing us to determine the ground states of the projected Hamiltonians, $\{\ket{\psi_\mathrm{GS}^k}\}$, via exact diagonalization.
These ground states are used to classically determine the average occupation of each orbital, i.e.
\begin{align*}
    \langle n_{i,\sigma} \rangle = \frac{1}{K}\sum_{k=1}^K \bra{\psi_\mathrm{GS}^k} n_{i,\sigma}  \ket{\psi_\mathrm{GS}^k}.
\end{align*}
To correct the faulty bitstrings from $S_X$, the number of bits that need to be flipped in order for the bitstring to be in the correct symmetry sector is determined.
For example, if a faulty bitstring has one more occupied orbital in the spin-up sector than needed, one of the occupied orbitals is chosen at random to be flipped with a certain probability $p$, that depends on the the difference between the orbital's occupation in the faulty bitstring and the orbital's average occupation $\langle n_{i,\sigma} \rangle$.
If the two occupations are similar, the bit is flipped with a low probability, and if the difference is close to one the bit is flipped with a high probability.
For further details see Ref.~\cite{RobledoMoreno2024} and Appendix~\ref{app:sccr}.
After having gone through all faulty bitstrings, some bitstrings out of $S_X$ may now lie in the correct symmetry sector.
If so, we remove them from $S_X$ and add them to $S_N$, re-evaluate $\langle n_{i,\sigma} \rangle$, and the procedure repeats until no faulty bitstrings are left in $S_X$.

\paragraph{\Ac{SD}~\cite{Kanno2023}:}
After sampling the optimized circuit and subsequently recovering faulty bitstrings, we use \ac{SD}, introduced in Ref.~\cite{Kanno2023}, to estimate the \ac{GSE}.
Similarly to solving the projected Hamiltonian during \ac{SCCR}, we construct a subspace Hamiltonian from all correct as well as recovered bitstrings.
This subspace Hamiltonian's \ac{GSE} can be calculated efficiently classically and yields an upper bound on the one for $H$.
We use this method for the final estimation of the variational \ac{GSE}.

\section{Results\label{sec:results}}
In this section we present our results in four steps.
We first present the successful parameter optimization on quantum hardware, yielding a classically tractable subspace.
From this, we estimate the \acp{GSE} by classically post-processing the results obtained from quantum hardware.
This allows us to determine the \acp{GSE} to well within chemical accuracy, i.e., to within 1.5\,mHa, of the exact solution of the active space Hamiltonians.
We then integrate the quantum hardware results into a quantum chemical workflow to arrive at the chemical property of interest, which is the difference in Gibbs free energy between the \ac{LS} and \ac{IS} states of Fe(III)-NTA in aqueous solution, and compare this to classical results.
Finally, we establish that our work covers one of the most complex quantum chemistry problems solved on quantum hardware to date by comparing the multi-reference diagnostics of the Fe(III)-NTA Hamiltonians to those of other molecular Hamiltonians that have been previously calculated on quantum hardware.

\paragraph{Parameter optimization on hardware:}
To train the parameters on quantum hardware, we choose the circuit ansatz with one layer.
Since each spin sector is acted upon by 10 Givens rotations within one layer, this amounts to 20 independently parameterized Givens rotations, and each Givens rotation can be decomposed using two 2QGEs and single-qubit gates.
The experiment itself was performed with a 10-ion chain of $^{40}$\textrm{Ca}$^+$ ions trapped in a linear Paul trap. We utilize an optical qubit encoded in \mbox{$\ket{0}=\ket{4 ^2\textrm{S}_{1/2}, m_J = -1/2}$} and \mbox{$\ket{1}=\ket{3 ^2\textrm{D}_{5/2}, m_J = -1/2}$} Zeeman sub-levels. An optical addressing system for \SI{729}{\nano\meter} laser light driving the transition between these qubit states allows for individual qubit control. In addition, ion-ion interaction through common motional modes of the trap provides all-to-all connectivity for two-qubit gates based on the M{\o}lmer-S{\o}rensen (MS) interaction~\cite{sorensen2000entanglement}. A more detailed description of the experimental setup is given in Refs.~\cite{pogorelov2021compact, postler2022demonstration, heussen2023strategies}. The native gate set available in this work consists of parameterized single-qubit rotations $R(\theta, \phi)$ in the $XY$ plane of the Bloch sphere, virtual $Z$ rotations, and fixed-angle $R_\mathrm{XX}(-\pi/2)$ rotations. Details on circuit transpilation to the hardware can be found in Appendix~\ref{app:qiskit}.
Together with the layer of CZ gates that entangles the spin sectors, the entire circuit then decomposes into 45 $R_\mathrm{XX}$, and 75 $R$ gates.

\begin{figure}[ht!]
    \centering
    \includegraphics[width=\linewidth]{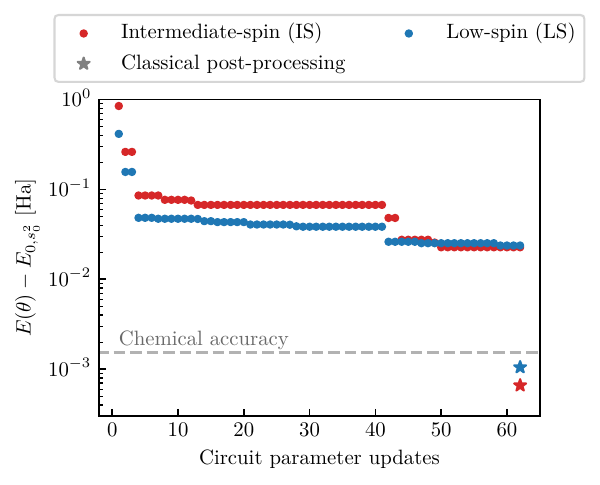}
    \caption{\label{fig:parameter_optimization}
        \textbf{Circuit parameter optimization on a trapped ion device.}
        Energy expectation values are calculated by measuring 600 times in the $Z^{\otimes N}$ basis, post-selecting and discarding erroneous outcomes, and estimating the diagonal terms of the penalized Hamiltonian $H$ (\autoref{eq:comp_base_estimate}).
        The dots show the estimation of energy differences between the trial and ground states, the dashed line shows the threshold below which chemical accuracy is achieved.
        After the final round of post-processing, this threshold is reached (shown by asterisks).
    }
\end{figure}
At each parameter update we obtain 600 single-shot measurements in the $Z^{\otimes N}$ basis from the ion trap.
Out of these 600 measurements, 30\,--\,40\,\% remain after post-selection, and those post-selected (correct) measurement outcomes are then used to estimate the energy, as seen in \autoref{fig:parameter_optimization}.
A new set of parameters is suggested by the \ac{NFT} optimizer \cite{Nakanishi2020, Qiskit}, and this process is repeated 60 times.
At the end of the optimization we are left with optimized circuits that put first upper bounds on the \acp{GSE}.
The differences to the exact ones, obtained via exact diagonalization of the entire active space Hamiltonian, are $31.6\pm1.3$\,mHa for the \ac{IS}, and $28.6\pm0.6$\,mHa for the \ac{LS} state of Fe(III)-NTA.
The uncertainty intervals correspond to $\sigma/\sqrt{N}$, where $\sigma$ is the samples' standard deviation and $N$ is the number of measurements (both after the post-selection process).

\paragraph{Estimating \acp{GSE}:}
After the optimization routine is finished, we sample the optimized circuits a final $10^4$ times in the $Z^{\otimes N}$ basis.
We again separate correct and erroneous bitstrings, and out of the $10^4$ measurements 30\,--\,40\,\% fulfill the post-selection criteria.
As outlined in the previous sections, we can make use of the \ac{SCCR} scheme to map the erroneous outcomes back to the correct symmetry sector~\cite{RobledoMoreno2024}.
Since this is a probabilistic procedure, we run the scheme 200 times for each of the two spin states and pick the sets yielding the lowest energies.
The final energy differences to the exact solutions are $0.66$\,mHa (\ac{IS}) and $1.05$\,mHa (\ac{LS}), meaning both systems were solved to within chemical accuracy of the exact solution on the active space.
Since the classical subroutines yield a classically efficient state representation, we calculate the states' uncertainties $\Delta H = \sqrt{\langle H^2\rangle -\langle H\rangle^2}$ to assess the quality of the results.
We obtain values of $15.7$\,mHa (\ac{LS}) and $4.5$\,mHa (\ac{IS}), indicating quantum states of very low variance and, thus, with high overlap to the actual respective ground states.

\paragraph{Energy difference between two spin states of Fe(III)-NTA:}
In order to ensure that the quantum computation actually enhances chemical research, the determined energies need to be embedded into a holistic quantum chemical workflow to finally arrive at thermodynamic properties that are of interest to chemists, e.g., by being measurable in chemical experiments. In this work, we use the \acp{GSE} from the previous paragraph to calculate the
difference in Gibbs free energy, $\Delta G^0$, between the LS and IS states of Fe(III)-NTA in aqueous solution.
We describe how $\Delta G^0$ is obtained in a classical post-processing step in Appendix~\ref{sec:qc_methods}.

The exact full numerical diagonalization of the Hamiltonian predicts the \ac{IS} state to be energetically more favourable than the \ac{LS} state by 33.7 mHa.
The quantum computation in combination with the \ac{SCCR} scheme predicts a $\Delta G^0$ of 34.1 mHa.
This difference of only 0.4 mHa is well within chemical accuracy.
Since the quantum processor is used to sample the bitstrings that are post-processed probabilistically by \ac{SCCR} and \ac{SD}, statistical uncertainties do not directly enter the classical determination of the \acp{GSE}, prohibiting access to meaningful uncertainty intervals.

A comparison with commonly used approximate classical methods is shown in \autoref{tab:Gibbs} revealing that also \ac{CCSD} yields a result that is within chemical accuracy. This is not unexpected, since single and double excitations are sufficient to cover almost all possible electron configurations within the rather small active space that was selected in this work. With \ac{MP2}, on the other hand, a significantly less accurate result is obtained, even predicting an incorrect energetic ordering of the two spin states. However, \ac{MP2}
still significantly improves upon \ac{HF}. The latter corresponds to the initial state that is prepared on the quantum hardware. We note that all contributions to $\Delta G^0$ besides the \acp{GSE} are independent of the method that is used to solve the active space Hamiltonian. Thus, the errors in $\Delta G^0$ can be attributed entirely to the errors in the solution of the active space Hamiltonian.

\begin{table}[]
\begin{tabular}{lr}
\toprule
Method                 & $\Delta G^0$ [mHa]  \\\midrule
Exact  (\textbf{c})& 33.7    \\[0.4em]
\ac{HF} (\textbf{c}) & -443.5 \\
MP2 (\textbf{c})         & $-0.8$    \\
CCSD (\textbf{c})        & 33.9    \\
Quantum algorithm (\textbf{q}, this work)       & 34.1 \\ \bottomrule

\end{tabular}
\caption{Difference in Gibbs free energy between the LS and IS states of Fe(III)-NTA in aqueous solution, $\Delta G^0$, using
classical (\textbf{c}) and quantum computing (\textbf{q}) methods for solving the respective active space Hamiltonian.
The \ac{HF} value also corresponds to the initial state prepared on the quantum processor.
The results from the quantum algorithm include the classical \ac{SCCR} and \ac{SD} post-processing steps.}\label{tab:Gibbs}
\end{table}

\paragraph{Multi-reference diagnostics of Fe(III)-NTA and comparison to other systems:}
\begin{table*}[ht!]
\begin{tabular}{llllllll}
\toprule
Molecular system                & Qubits      & Electrons  & Orbitals   & Source                                    & $D_1$ & $T_1$ & $Z_{S(1)}$     \\ \midrule
\textbf{Fe(III)-NTA LS}      & \textbf{10} & \textbf{5} & \textbf{5} & \textbf{This work}                        & \textbf{0.166} & \textbf{0.076} & \textbf{0.207} \\
\textbf{Fe(III)-NTA IS}      & \textbf{10} & \textbf{5} & \textbf{5} & \textbf{This work}                        & \textbf{0.029} & \textbf{0.013} & \textbf{0.299} \\
Diazene TS2                  & 10          & 12         & 10         & \cite{Arute2020}         & 0.235          & 0.070          & 0.217          \\
{[}Fe$_4$S$_4$(SCH$_3$)$_4${]}$^{2-}$  & 77          & 54         & 36         & \cite{RobledoMoreno2024} & 0.176          & 0.046          & 0.170           \\
NOR intermediate A           & 8           & 4          & 4          & \cite{Ollitrault2024}     & 0.171          & 0.085          & 0.164          \\
{[}Fe$_2$S$_2$(SCH$_3$)$_4${]}$^{2-}$ & 45          & 30         & 20         & \cite{RobledoMoreno2024} & 0.164          & 0.040          & 0.122         \\
NOR intermediate B           & 8           & 4          & 4          & \cite{Ollitrault2024}     & 0.161          & 0.081          & 0.177          \\
Li$_2$O                      & 12          & 8          & 12         & \cite{Zhao2023}          & 0.068          & 0.036          & 0.106           \\
Sulfonium triplet            & 6           & 6          & 6          & \cite{Motta2023EF}       & 0.034          & 0.018          & 0.080          \\
Diazene TS1                  & 10          & 12         & 10         & \cite{Arute2020}         & 0.033          & 0.012          & 0.110         \\
Cyclobutene conrotatory TS   & 6           & 6          & 6          & \cite{obrien2023}        & 0.026          & 0.011          & 0.172          \\
N$_2$                        & 58          & 14         & 18         & \cite{RobledoMoreno2024} & 0.025          & 0.010          & 0.076           \\
Diazene (cis)                & 10          & 12         & 10         & \cite{Arute2020}         & 0.025          & 0.007          & 0.114          \\
Cyclobutene conrotatory TS   & 10          & 10         & 10         & \cite{obrien2023}        & 0.024          & 0.009          & 0.112          \\
Sulfonium singlet            & 6           & 6          & 6          & \cite{Motta2023EF}       & 0.023          & 0.010          & 0.079          \\
H$_{12}$ (chain)             & 12          & 12         & 12         & \cite{Arute2020}         & 0.020          & 0.007          & 0.144          \\
Diazene (trans)              & 10          & 12         & 10         & \cite{Arute2020}         & 0.018          & 0.006          & 0.112           \\
H$_{10}$ (chain)             & 10          & 10         & 10         & \cite{Arute2020}         & 0.017          & 0.007          & 0.139          \\
H$_2$O                       & 6           & 8          & 6          & \cite{Zhao2023}          & 0.016          & 0.006          & 0.066          \\
H$_8$ (chain)                & 8           & 8          & 8          & \cite{Arute2020}         & 0.014          & 0.006          & 0.134          \\
H$_2$O                       & 5           & 6          & 5          & \cite{Eddins2022EF}      & 0.014          & 0.006          & 0.070         \\
H$_6$ (chain)                & 6           & 6          & 6          & \cite{Arute2020}         & 0.011          & 0.005          & 0.126          \\
Butadiene                    & 6           & 6          & 6          & \cite{obrien2023}        & 0.011          & 0.006          & 0.190          \\
Diels-Alder TS & 8           & 8          & 8          & \cite{Liepuoniute2024EF} & 0.010          & 0.004          & 0.146          \\
Cyclopentadiene   & 6           & 6          & 6          & \cite{Liepuoniute2024EF} & 0.009          & 0.004          & 0.123          \\
Butadiene                    & 10          & 10         & 10         & \cite{obrien2023}        & 0.008          & 0.004          & 0.123          \\
Cyclobutene                  & 6           & 6          & 6          & \cite{obrien2023}        & 0.008          & 0.003          & 0.077          \\
Cyclobutene                  & 10          & 10         & 10         & \cite{obrien2023}        & 0.007          & 0.003          & 0.059          \\
Diels-Alder TS & 6           & 6          & 6          & \cite{Liepuoniute2024EF} & 0.006          & 0.003          & 0.186          \\
BeH$_2$                      & 6           & 4          & 4          & \cite{Kandala2017}       & 0.003          & 0.002          & 0.065          \\
Li$_4$                       & 8           & 4          & 4          & \cite{sun2024evaluating} & 0.001          & 0.001          & 0.031          \\
{[}CH$_3$· – H – OH{]} TS      & 6           & 3          & 3          & \cite{khan2023}          & 0.001          & 0.001          & 0.002          \\
CH$_3$·                      & 6           & 3          & 3          & \cite{khan2023}          & 0.001          & 0.000          & 0.005          \\
LiH                          & 6           & 2          & 3          & \cite{guo2024}           & 0.000          & 0.000          & 0.034           \\
CH$_4$                       & 6           & 2          & 3          & \cite{khan2023}          & 0.000          & 0.000          & 0.011          \\
F$_2$                        & 12          & 10         & 6          & \cite{guo2024}           & 0.000          & 0.000          & 0.057          \\
H$_4$ (square)               & 8           & 4          & 4          & \cite{huggins2022AFQMC}  & 0.000          & 0.000          & 0.476          \\
N$_2$                        & 12          & 6          & 6          & \cite{huggins2022AFQMC}  & 0.000          & 0.000          & 0.096          \\
OH·                          & 6           & 3          & 3          & \cite{khan2023}          & 0.000          & 0.000          & 0.271          \\
H$_2$O                       & 6           & 2          & 3          & \cite{khan2023}          & 0.000          & 0.000          & 0.005
\\
\bottomrule
\end{tabular}
\caption{$D_1$, $T_1$ and $Z_{S(1)}$ multi-reference diagnostics for the active spaces of different molecular systems calculated on quantum hardware. The systems from the literature are sorted by decreasing $D_1$ diagnostic. The number of utilized qubits is shown together with the number of electrons and spatial orbitals contained in the selected active spaces, which were reconstructed as described in the respective source.
TS refers to transition state and NOR stands for nitric oxide reductase. Several molecular geometries are not provided by the respective source. The structure of {[}CH$_3$· – H – OH{]} TS was taken from~\cite{espinosa2015}. The structures of the NOR intermediates, butadiene, cyclobutene and its conrotatory TS were obtained from the respective authors. The structures of sulfonium, CH$_3$·, CH$_4$ and OH· were optimized manually. For further information see Appendix~\ref{sec:qc_methods}.}\label{tab:complexity_chemistry}
\end{table*}

We estimate the multi-reference character of the selected active space Hamiltonians by means of the $D_1$ and $T_1$ diagnostics from \ac{CCSD} calculations~\cite{Jansen1998D1,Lee1989T1}. These diagnostics are commonly used in quantum chemistry to determine whether or not a single-reference wavefunction can describe a system well. As empirical thresholds for entire molecules, a $D_1$ diagnostic exceeding 0.05 and a $T_1$ diagnostic exceeding 0.02 suggest the usage of multi-reference methods. Here, we calculate these diagnostics for several active space Hamiltonians.
We also include the $Z_{S(1)}$ multi-reference diagnostic as an additional measure that does not rely on \ac{CCSD} but instead is based on single-orbital entropies obtained with \ac{DMRG} theory~\cite{Stein2017ZS1}. Further information can be found in Appendix~\ref{sec:qc_methods}.

\autoref{tab:complexity_chemistry} presents these diagnostics for the two selected spin states of Fe(III)-NTA as well as for other molecular systems previously calculated on quantum hardware, which utilized \acp{VQE} comprising at least five qubits. The number of electrons and spatial orbitals considered in the respective calculations is also given. For studies on chemical reactions, we restrict our analysis to reactants, \acp{TS}, and products, and for dissociation curves we only consider the geometry with the lowest energy.
Six systems exhibit $D_1$ diagnostics that significantly exceed the empirical threshold of 0.05, thus rendering them multi-reference systems: the out-of-plane \ac{TS} of the diazene isomerization denoted as ``Diazene TS2'' in \autoref{tab:complexity_chemistry}~\cite{Arute2020}, the iron-sulfur clusters ${[}$Fe$_2$S$_2$(SCH$_3$)$_4${]}$^{2-}$ and ${[}$Fe$_4$S$_4$(SCH$_3$)$_4${]}$^{2-}$~\cite{RobledoMoreno2024}, the two \ac{NOR} intermediates~\cite{Ollitrault2024}, and the \ac{LS} state of Fe(III)-NTA from this work. The \ac{TS} of diazene shows the largest $D_1$ diagnostic (0.24), and the remaining five systems, all containing at least one transition metal atom, exhibit $D_1$ diagnostics of around 0.17, which are more than twice as large as the next highest values. Similarly, the $T_1$ diagnostics of all of the aforementioned systems also clearly exceed the empirical threshold of 0.02. For the large-sized active spaces of the iron-sulfur clusters, the $T_1$ diagnostics are small compared to their $D_1$ diagnostics, which is likely due to the known dependence of $T_1$ on the problem size~\cite{Jansen1998D1}. Furthermore, it is noted that we were not able to determine the $D_1$ and $T_1$ diagnostics for the broken-symmetry \ac{LS} states of the two iron-sulfur clusters  due to non-convergence of the underlying \ac{CCSD} calculations. Instead, we determined these diagnostics for the respective \ac{HS} states, which likely underestimate the hypothetical diagnostics for the broken-symmetry \ac{LS} states.

The $Z_{S(1)}$ diagnostics paint a less clear picture that does not match the trend observed in the $D_1$ or $T_1$ diagnostics in many cases. For the \ac{LS} state of Fe(III)-NTA, the large $Z_{S(1)}$ diagnostic of 0.21 qualitatively agrees with the large $D_1$ and $T_1$ diagnostics, supporting its classification as a multi-reference system. For the \ac{IS} state, the $Z_{S(1)}$ diagnostic even amounts to 0.30, while the $D_1$ and $T_1$ diagnostics are below the respective empirical thresholds (amounting to 0.03 and 0.01, respectively).
In comparison to the other molecular systems investigated in \autoref{tab:complexity_chemistry}, both spin states of Fe(III)-NTA exhibit among the largest $Z_{S(1)}$ diagnostics.
It should, however, be noted that the $Z_{S(1)}$ diagnostics are generally less reliable for cases where the number of electrons and spatial orbitals are unequal~\cite{Stein2017ZS1} such as the \ac{TS} of diazene and the iron-sulfur clusters.
Overall, the large $D_1$, $T_1$, and $Z_{S(1)}$ diagnostics of the \ac{LS} state of Fe(III)-NTA underscore its strong multi-reference character, making it a valuable test case for quantum computations.

Finally, we compare the accuracy we reach for Fe(III)-NTA using quantum hardware to that obtained for the other systems with similarly strong multi-reference character, i.e. with similarly strong correlations.
As discussed above, we achieve chemical accuracy for the \acp{GSE} of both the \ac{LS} and \ac{IS} state of Fe(III)-NTA (deviations from the exact solutions of $1.05$\,mHa and $0.66$\,mHa, respectively).
In the case of the \ac{NOR} intermediates, the electrostatic interaction energy, which is only one contribution to the \ac{GSE}, was determined on quantum hardware to within chemical accuracy, namely to within $0.3$\,mHa for intermediate A and $0.6$\,mHa for intermediate B of the exact solution, with a quantum circuit optimized by classical simulations~\cite{Ollitrault2024}. For the active spaces of ${[}$Fe$_2$S$_2$(SCH$_3$)$_4${]}$^{2-}$ and ${[}$Fe$_4$S$_4$(SCH$_3$)$_4${]}$^{2-}$, due to their larger size, the obtained \acp{GSE} deviate from the \ac{DMRG} solution roughly by $150$\,mHa and $650$\,mHa, respectively~\cite{RobledoMoreno2024}. It is noted, however, that the study achieves significantly greater accuracy for the \acp{GSE} of these systems using energy-variance extrapolation, although precise values and error estimates are not given. The system with the largest $D_1$ diagnostic, the out-of-plane \ac{TS} of the diazene isomerization, was solved at the \ac{HF} level of theory, which does neither account for the multi-reference character of the molecule nor for electron correlation at all. The deviation from the classically obtained \ac{HF} energy is roughly $7$\,mHa~\cite{Arute2020}.

\section{Conclusion\label{sec:discussion}}
In this work, we used a trapped ion quantum computer to calculate energies of active space Hamiltonians of the industrially relevant Fe(III)-NTA metal chelate complex to within chemical accuracy. The results were integrated into a typical quantum chemical workflow to arrive at the Gibbs free energy in aqueous solution, which is one of the most important thermodynamic properties in chemical research. The two considered spin states of Fe(III)-NTA are amongst the most complex molecular problems studied using quantum hardware to date, according to their multi-reference diagnostics.
By finding the ground states of the respective Hamiltonians, we demonstrated that a variational quantum algorithm can be trained on quantum hardware to provide a classically tractable subspace, in which accurate approximations to the ground state and its energy can be found via classical post-processing. This mitigates some of the main challenges that variational quantum algorithms face in their conventional form: the preparation of the subspace was achieved with a cost function that required significantly fewer measurements than estimates of energy expectation values or their derivatives. At the same time, it enabled an efficient post-selection of the measurement results to filter out erroneous runs. Using this algorithm, our work achieved chemical accuracy for the energetic separation between the low- and intermediate-spin state of Fe(III)-NTA.
While practical quantum advantage still remains out of reach, we believe that this work nevertheless makes an important contribution to solving industrially relevant and complex quantum chemistry problems on quantum hardware.

\section*{Data availability}
The datasets used and analyzed in this paper are available from the corresponding authors upon reasonable request.
\vspace{-\baselineskip}

\section*{Code availability}
The code used to generate the data for this paper is available from the corresponding authors upon reasonable request.
\vspace{-\baselineskip}

\section*{Acknowledgements}
This work was supported by the German Federal Ministry of Education and Research (BMBF)
within the
funding program “Quantum technologies — From basic
research to market” in the joint project MANIQU (Grant
Nos. 13N15577, 13N15578 and 13N15575), and the Munich Quantum Valley, which is supported by the Bavarian state government with funds from the Hightech Agenda Bayern Plus.
The Innsbruck team acknowledges support by the European Research Council (ERC, QUDITS, 101080086), and the European Union’s Horizon Europe research and innovation programme under grant agreement No 101114305 (``MILLENION-SGA1'' EU Project) and Grant Agreement Number 101046968 (BRISQ). Views and opinions expressed are however those of the author(s) only and do not necessarily reflect those of the European Union or the European Research Council Executive Agency. Neither the European Union nor the granting authority can be held responsible for them. We also acknowledge support by the Austrian Science Fund (FWF Grant-DOI 10.55776/F71) (SFB BeyondC), the Austrian Research Promotion Agency under Contracts Number 897481 (HPQC), the Institut für Quanteninformation GmbH.

\vspace{\baselineskip}
\section*{Competing Interests}
T. M. is connected to Alpine Quantum Technologies GmbH, a commercially oriented quantum computing company.

\section*{Author contributions}
L. N. designed and implemented the variational circuit, parameter optimization, post-processing, and error mitigation process.
A. G. implemented the circuit read-out (including alternative methods) and guided the post-processing and error mitigation process.
L. H. prepared the molecular Hamiltonians and calculated the diagnostics and classical solutions.
L. M. conducted preliminary numerics and tested alternative methodologies.
R. F. prepared and implemented the measurements and maintained the apparatus.
A. S. maintained the experimental apparatus.
Ch. M. built the apparatus and supervised the experimental team.
T. E., N. S. and S. W. implemented classical post-processing techniques.
T. M. supervised the experimental team and acquired funding.
M. K. selected the quantum chemistry problem, supervised the quantum chemistry team, and acquired funding.
M. J. H. supervised the entire project and acquired funding.

\begin{acronym}

\acro{BK}{Bravyi-Kitaev}

\acro{CASCI}{complete active space configuration interaction}
\acro{CASSCF}{complete active space self-consistent field}
\acro{CCSD}{coupled-cluster singles and doubles}

\acro{DMRG}{density matrix renormalization group}
\acro{DFT}{density functional theory}

\acro{GSE}{ground-state energy}
\acroplural{GSE}[GSEs]{ground-state energies}

\acro{HF}{Hartree-Fock}
\acro{HS}{high-spin}

\acro{IS}{intermediate-spin}

\acro{LS}{low-spin}

\acro{MP2}{second-order M{\o}ller-Plesset perturbation theory}

\acro{NFT}{Nakanishi-Fujii-Todo}
\acro{NISQ}{noisy intermediate-scale quantum}
\acro{NP}{number-preserving}
\acro{NOR}{nitric oxide reductase}

\acro{QNP}{quantum-number-preserving}
\acro{QUBO}{quadratic unconstrained binary optimization}

\acro{SCCR}{self-consistent configuration recovery}
\acro{SPAM}{state preparation and measurement}
\acro{SD}{subsystem diagonalization}

\acro{TS}{transition state}
\acroplural{TS}[TS]{transition states}

\acro{VQA}{variational quantum algorithm}
\acro{VQE}{variational quantum eigensolver}

\acro{ZNE}{zero-noise-extrapolation}

\end{acronym}

\bibliography{literature.bib}

\appendix

\section{Circuit Ansatz\label{subsec:circuit_ansatz}}
Having chosen a fermion-to-qubit mapping and cost function leaves us with choosing a suitable circuit ansatz.
In accordance with the reasoning applied to enforcing the total spin squared through the cost function, we seek a circuit ansatz that conserves particle numbers.
There are numerous \ac{NP} circuits, including the UCCSD~\cite{Bartlett1989, Taube2006, Filip2020}, ADAPT-VQE~\cite{Grimsley2019}, or Hamiltonian Variational Ansatz~\cite{Wecker2015,Wiersema2020}.
We refer to Ref.~\cite{Tilly2022} for a comprehensive overview of ansaetze.
Many of the \ac{NP} ones, however, require the implementation of exponentials of Hamiltonian terms that are mapped to Pauli strings with Pauli weight $\mathcal{O}(N)$ in the Jordan-Wigner transform.
While one can circumvent this by employing local fermionic encodings where additional qubits are introduced to reduce the Pauli weight of the terms in the Hamiltonian, we focus on easily implementable circuits with low qubit numbers.
Hence, we choose an \ac{NP} circuit consisting of two-local operations.

While there are multiple \ac{NP} two-local operations one could choose~\cite{Egger2019,McKay2016,Sagastizabal2019}, we settle on parameterized Givens rotations~\cite{Anselmetti2021},
\begin{align} \label{eq:givens}
    G(2\theta) =
    \begin{bmatrix}
        &&&\\[-1em]
        1 &  &  & \\
        & \cos \theta & \sin \theta & \\
        & -\sin \theta & \cos \theta & \\
        &  &  & 1\\
    \end{bmatrix}
\end{align}
that can be implemented using two 2QGEs.
The reason for this choice is quite simple: The Hamiltonian in \autoref{eq:electronic_structure_ham} is real, making the Hamiltonian's eigenvectors real.
The circuit therefore only needs to explore the real part of the Hilbert space to find the target state, which is achieved with Givens rotations.

Overall, we opted for a layered variational ansatz, where each layer consists of two parts: a set of Givens rotations and CZ gates.
Since the particle number is conserved in each spin sector, we apply sets of Givens rotations within the subsectors separately.
This leads to the spin sectors forming a product state in the first layer, and we subsequently apply CZ gates between the spin sectors to allow for entanglement without particle exchange.
One layer of the variational ansatz is depicted in the bottom of \autoref{fig:circuit_ansatz}.

In the end, the entire circuit is made up of three parts: the initialization, the evolution under the \ac{NP} circuit, and the measurement, as shown in the top part of \autoref{fig:circuit_ansatz}.
Here, we choose the initialization step to be the trivial preparation of the HF state, and the measurement to only consider single-qubit readout in the $Z^{\otimes N}$ basis.

\section{Self-Consistent Configuration Recovery\label{app:sccr}}
\begin{figure}[h!]
    \centering
    \includegraphics[width=\linewidth]{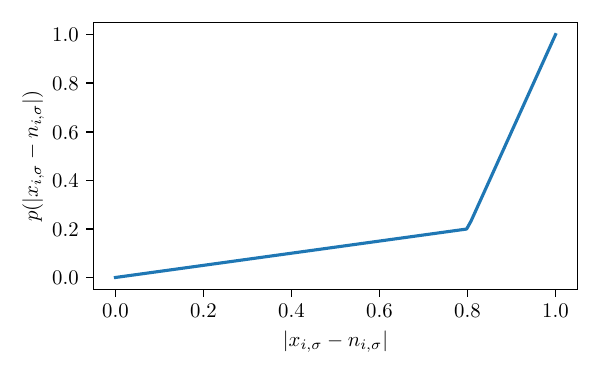}
    \caption{
    \textbf{Probability weight.}
    Probability weight of flipping a faulty bit dependent on the corresponding orbital's average occupation $\langle n_{i,\sigma}\rangle$, obtained from the set of correct(ed) samples $S_N$.}
    \label{fig:ReLU}
\end{figure}

As explained in Ref.~\cite{RobledoMoreno2024}, the SCCR scheme flips certain bits in a bitstring probabilistically.
The probability of flipping a bit is dependent on the absolute difference between the corresponding orbital's average occupation $\langle n_{i,\sigma}\rangle$, obtained from the set of correct(ed) samples $S_N$, and the orbital's occupation in the faulty bitstring, $x_{i,\sigma}$.
Just as suggested in Ref.~\cite{RobledoMoreno2024}, we determine the probability weight of flipping the bit using the modified rectified linear unit (ReLU) function $p(|x_{i,\sigma}-n_{i,\sigma}|)$,
\begin{align}
    p(y) = \begin{cases}
        \delta \frac{y}{h},\quad &y\leq h,\\
        \delta + (1-\delta)\frac{y-h}{1-h},& y>h.
    \end{cases}
\end{align}
We choose the hyperparameters to be $h = 0.8$ and $\delta = 0.2$, which results in the ReLu function shown in \autoref{fig:ReLU}.

\section{Circuit transpilation} \label{app:qiskit}
Circuits within this work were submitted to the experimental device via our custom modification of the Qiskit backend by Alpine Quantum Technologies (AQT)~\cite{aqt_provider}. The customized provider has a transpilation stage to adapt a circuit for running on the experimental setup to enhance performance. The transpilation procedure consists of several stages:
\begin{enumerate}
    \item Using Qiskit default transpiler with optimization level 3~\cite{qiskit_transpiler} to simplify the circuit and convert all the gates to the setup's native gates set: \mbox{$R(\theta, \phi) = \exp(-i\frac{\theta}{2}(X \cos \phi + Y \sin \phi))$} and $R_{XX}(-\pi/2)$.
    \item Minimizing the number of single-qubit gates by combining them and commuting gates through $R_{XX}$ gates where possible according to~\cite{maslov2017}.
    \item Discarding all $R(\theta, \phi)$ gates with $|\theta| < 10^{-3}$. This is motivated by the fact that the infidelity of a physical single-qubit gate in the setup is on the order of $10^{-4}$, so performing a gate physically will not be better than skipping it.
\end{enumerate}
Running this transpilation procedure yields a circuit with fewer gates than the initial circuit, and thus a higher fidelity measurement result.

\section{Quantum Chemical Methods}
\label{sec:qc_methods}
Classical quantum chemical calculations were conducted using \texttt{PySCF}~\cite{Sun2020pyscf}. Computationally optimized structures for the \ac{IS} and \ac{LS} states of Fe(III)-NTA are taken from Ref.~\cite{Hehn2024}. Basis sets of quadruple-\textzeta{} valence quality with two sets of polarization functions (def2-QZVPP)~\cite{weigend2003a,Weigend2005def2bases} were used for all classical calculations as well as for the generation of the active space Hamiltonians.
The selected (5,5) active space of Fe(III)-NTA comprises five molecular orbitals with predominantly iron $3d$ character that are occupied by five electrons in such a way that the occupation adheres to the selected spin state.
The active space Hamiltonians for the quantum computations are then calculated in the basis that is given by the molecular orbitals of a respective \ac{CASSCF} calculation.

\paragraph{Multi-reference diagnostics}
The multi-reference character of an active space Hamiltonian can be estimated by means of multi-reference diagnostics as already stated in \autoref{sec:results}.
The $T_1$ and $D_1$ diagnostics are generally determined from the single excitation amplitudes obtained from a \ac{CCSD} calculation, wherein the
$T_1$ diagnostic is defined as their Frobenius norm and the $D_1$ diagnostic as their matrix 2-norm~\cite{Golub2013}.
Furthermore, we calculate the $Z_{S(1)}$ diagnostic that is based on single-orbital entropies from \ac{DMRG} theory ~\cite{Stein2017ZS1}. We run \ac{DMRG} calculations using \texttt{PySCF} in combination with \texttt{Block2}~\cite{Zhai2023block2} and a maximum bond dimension of 500.

The molecular geometries of several systems in
\autoref{tab:complexity_chemistry} are not provided by the respective source. In such cases, we either took the geometry from another source and cited it in the caption of \autoref{tab:complexity_chemistry}, or we optimized the geometry using {\texttt{TURBOMOLE}}~\cite{AHLRICHS1989TM,Treutler1995TMDFT,Furche2014TMWires}. If provided, we used the methods and basis sets given by the source. If not provided, we optimized the structures using the B3LYP hybrid density functional~\cite{Becke1993B3LYP} in combination
with basis sets of triple-\textzeta{} valence quality with one set of polarization functions (def2-TZVP)~\cite{weigend2003a,Weigend2005def2bases}.
The reconstruction of the active spaces follows the sources as closely as possible as well. In case details are missing, we selected the active space based on orbitals from a converged restricted or restricted open-shell \ac{HF} calculation.

\paragraph{Calculation of Gibbs free energies in aqueous solution}
In a typical industrial quantum chemical workflow the \ac{GSE} obtained by solving the respective active space Hamiltonian is further processed to arrive at the Gibbs free energy $G^0$. This is done by adding thermodynamic and solvation contributions to the \ac{GSE}~\cite{Jensen2015}. We used the same approach as described for Fe(III)-NTA in Ref.~\cite{Hehn2024}: statistical thermodynamic contributions were included using the rigid-rotor and harmonic-oscillator (RRHO) approximations for a temperature of 25°C. Solvation in water was done using the COSMO-RS method~\cite{Klamt1995,Klamt1998,Klamt2001} (version 2018 of the {\texttt{COSMOtherm}} software~\cite{COSMOtherm, Eckert2002}) together with the fine parametrization based on BP86/def2-TZVPD~\cite{Becke1988BP86,rappoport2010property} assuming an infinite dilution.

After obtaining $G^0$ for each of the two selected spin states of Fe(III)-NTA, the difference in Gibbs free energy, $\Delta G^0$, is simply the result of subtracting $G^0$(\ac{IS}) from $G^0$(\ac{LS}).

\end{document}